%% file: main.tex
\documentclass{ametsocV6.1}

\usepackage[]{babel}
\usepackage[utf8]{inputenc}
\usepackage{float}
\usepackage{amsmath,amssymb}
\usepackage{bm}
\usepackage{graphicx}
\usepackage{url}
\usepackage{caption}
\usepackage{subcaption}

\usepackage{hyperref}
\hypersetup{
    colorlinks=true,
    linkcolor=blue,
    filecolor=magenta,      
    urlcolor=cyan,
    citecolor=black,
    pdftitle={Overleaf Example},
    pdfpagemode=FullScreen,
    }

\usepackage[normalem]{ulem} 

\title{Application of deep learning to the estimation of normalization coefficients in diffusion-based covariance models}

\authors{
Folke Skrunes\aff{a},
Mayeul Destouches\aff{b,a}\correspondingauthor{Mayeul Destouches, mayeul.destouches@metoffice.gov.uk}
Anthony T. Weaver\aff{a},
Guillaume Coulaud\aff{a},
Olivier Goux\aff{a},
Corentin Lapeyre\aff{a}
}

\affiliation{
\aff{a}{CERFACS / CECI CNRS, UMR 5318 , Toulouse, France},
\aff{b}{Met Office, Exeter, United Kingdom}
}

\abstract{
Variational data assimilation in ocean models depends on the ability to model general correlation operators in the presence of coastlines. Grid-point filters based on diffusion operators are widely used for this purpose, but come with a computational bottleneck -- the costly estimation of normalization factors for every model grid point. 
In this paper, we show that a simple convolutional neural network can effectively learn these normalization factors with better accuracy than the current operational methods. 
Our network is tested with a two-dimensional diffusion operator from the NEMOVAR ocean data assimilation system, applied to a global ocean grid with approximately one degree horizontal resolution.
The network is trained on exact normalization factors estimated by a brute-force method. 
Knowing that convolutional networks can only model translation-equivariant functions, we ensure that the normalization estimation problem is indeed translation-equivariant. 
Specifically, we show how the number of inputs of this problem can be reduced while preserving translation equivariance.
Adding the distance to the coastline as an input channel is found to improve the performance of the network around coastlines.
Extensions to three-dimensional diffusion and to higher horizontal resolutions are discussed. 
Removing the computational bottleneck associated with normalization opens the way to using adaptive correlation models for operational ocean data assimilation.
The code for this work is publicly available at
\url{https://github.com/FolkeKS/DL-normalization/tree/core-features}
}

\begin{document}

\maketitle

%
%
%
%
%
%

%

\input{./1-Intro.tex}
\input{./3-DL.tex}
\input{./4-Results.tex}
\input{./5-Discussion.tex}

\acknowledgments
The authors wish to thank Andrea Piacentini for his help in setting up and running the NEMOVAR experiments.

%
%
\datastatement
The code for this work is publicly available at
\url{https://github.com/FolkeKS/DL-normalization/tree/core-features} and the associated data set at \url{https://kaggle.com/datasets/e6e1e745161300b59a7ecf7695f02fedd5ab6f5853067004c127029d15fde68e}

%

\appendix
\input{./6-Appendix.tex}

 \bibliographystyle{ametsocV6}
 \bibliography{references}

\end{document}

%% file: 1-Intro.tex
\section{Introduction}
\label{sec:introduction}

Error covariance matrices are essential for determining how different pieces of information are weighted in a data assimilation (DA) problem.
For large problems, such as those typically encountered in atmospheric and ocean DA applications, it is impossible to represent them as explicit full-rank matrices.
Instead, they must be {\em modeled}, either as (localized) reduced-rank matrices or
in terms of parameterized covariance functions that have a tractable number of tunable parameters and that are computationally efficient when applied as the kernel of a covariance operator \citep{bannister2008reviewb}.
Covariance operators are commonly used in variational DA where only matrix-vector products are required for the iterative algorithms used to obtain the solution.
In this article, we consider covariance matrices belonging to this latter (operator) class.
Specifically, we focus on covariance operators that can be formulated as a sequence of differential operators and interpreted as the solution of a diffusion equation \citep{weaver_correlation_2001, mirouze_weaver_2010, weaver_diffusion_2013}.
When the diffusion equation is solved implicitly with an Euler-backward step, diffusion operators represent covariance operators whose smoothing kernels are from the Mat\'ern class \citep{guttorp06,lindgren_explicit_2011}.

Diffusion-based covariance operators are convenient in DA applications involving
domains with complex boundary geometry and requiring inhomogeneous and anisotropic smoothing tensors.
This is the case for background-error covariances in ocean DA \citep{weaver_correlation_2001, weaver2020evaluation}.
Specifically, and especially within the context of this article, these operators are needed to model {\em correlation} operators, which, by definition, correspond to covariance matrices whose diagonal elements (variances) are all equal to one. The symmetric property of the covariance matrix is maintained by pre- and post-multiplying the diffusion-based correlation matrix by the diagonal matrix of standard deviations.
By factoring the covariance operator this way, the correlations and variances can be controlled independently.

In order to isolate the correlation operator, diffusion-based covariance matrices must be normalized by their intrinsic variances.
We refer to the diagonal matrix whose elements are the inverse of the intrinsic variances as the matrix of {\em normalization factors}.
The normalization factors act to compensate the attenuation of the covariance amplitude that occurs from a diffusive process.
As with the actual variances, the normalization factors can be applied symmetrically through pre- and post-multiplication by a diagonal matrix that contains the
square root of the normalization factors.
The normalization factors have units of length, area or volume depending on whether the covariance operator is applied in one, two or three dimensions.

Except in idealized or specific cases, for which we can resort to well-known analytical expressions \citep{guttorp06}, the normalization factors are not known to high accuracy at all grid points.
Their values depend on the diffusion model parameters, such as the smoothing tensor; on the boundary geometry, such as coastlines in an ocean model; on the type of boundary conditions (Neumann, Dirichlet or mixed); and on the details of the numerical discretization scheme and computational solver.
Regardless of the domain and covariance model details, it is always possible to diagnose the {\em exact} normalization factor at any grid point using brute force, whereby the covariance matrix is applied to a canonical vector, which has a value of one at the grid point of interest and a value of zero at all other grid points.
However, this method is not practical for model configurations involving millions of grid points, especially if normalization factors have to be regularly recomputed, as is required in a cycled DA system when newly available estimates of the diffusion model parameters are applied on each cycle.
A randomization or Monte Carlo method is much more practical than the brute-force method as it can be used to generate sufficiently accurate normalization factors for any number of grid points at the cost of several thousand applications of the diffusion operator. 
Nevertheless, this is still too costly to be applied routinely ({\em i.e.}, on each DA cycle) to three-dimensional (3D) variables in a real-time operational environment where timeliness is critical.

Using a diffusion-based covariance model for a global ocean DA system, \cite{weaver2020evaluation} compared randomization with various other methods for approximating the normalization factors.
Their work showed that none of the other methods is accurate and robust enough to provide
a viable alternative to randomization.
As a compromise between allowing for adaptive covariances on the one hand, and keeping computational costs acceptable on the other, \cite{weaver2020evaluation} proposed a quasi-adaptive correlation model for 3D variables that relies on splitting the diffusion operator into a one-dimensional (1D) vertical component and a two-dimensional (2D) horizontal component (referred to by \cite{weaver2020evaluation} as a 2D$\times$1D formulation) and only allowing the 1D component to be adaptive.
Furthermore, they proposed a method whereby the normalization factors can be well approximated by a product of normalization factors computed separately for the 1D and 2D diffusion-model components.
This way, the normalization factors for the adaptive and inexpensive 1D component can be estimated ``on-the-fly'' to a high accuracy using the randomization or even brute-force method.
The normalization factors for the static and expensive 2D component applied in each model level need to be estimated only once, using the conventional randomization approach. 
In this article, we show how 2D normalization factors in the 2D$\times$1D formulation can be estimated accurately and cheaply using a machine-learning strategy, thus opening the way for the use of 2D adaptive correlations.

The article is organized as follows.
In Section~\ref{sec:deep-learning}, we show how the problem of estimating normalization factors for the diffusion operator can be transformed to make it suitable for a solution algorithm based on a Convolutional Neural Network (CNN).
Specific care is taken to ensure that the function to regress does indeed lie in the class of functions that can be approximated with any accuracy by CNNs.
The method has been applied to a 2D diffusion operator that forms the basis of a correlation model in a global ocean DA system.
Results are presented and discussed in Section~\ref{sec:results} where they are compared to those computed with a standard randomization method.
A summary and outlook are given in Section~\ref{sec:conclusion}.

%% file: 3-DL.tex
\section{Proposed method}
\label{sec:deep-learning}

We present here the machine learning approach used to estimate normalization factors for a diffusion-based covariance operator. For at least two reasons, the
natural choice for this problem is a Convolutional Neural Network (CNN).
First, the data are stored on a structured grid, where every point has neighboring grid points in all four directions.
This makes it very similar to an image-to-image translation exercise, where CNNs are known to perform well. 
The structured nature of the data allows us to use standard CNNs, instead of more general but less efficient Graph Neural Networks for instance. 
Second, the normalization problem is essentially local. 
While a normalization factor at a given location is, in theory, affected by the diffusion tensor at all points on the globe, in practice, it is mainly sensitive to the diffusion tensor at points in a small neighborhood. 
The impact of distant diffusion tensors decays roughly with the amplitude of the modeled correlation function. 
In DA applications, the modeled correlations are approximately Gaussian, with length scales that are short compared to the domain size. 
As a result, they decay to zero quite rapidly and the problem can effectively be considered as local.
This makes CNNs with even relatively small receptive fields good candidates for solving the normalization problem.

\subsection{Reducing the dimension of the problem}
\label{subsec:reducedim}

We now explain how the problem can be simplified by reducing the number of inputs required to train the network. 
No accuracy is lost in the process.
Our assumption is that reducing the number of inputs (a.k.a.\ channels) will make the problem easier to learn for a neural 
network, and will reduce the memory cost and execution time of both training and applying the network.

\begin{figure}[!h]
  \begin{center}
    \includegraphics[width=0.4\linewidth]{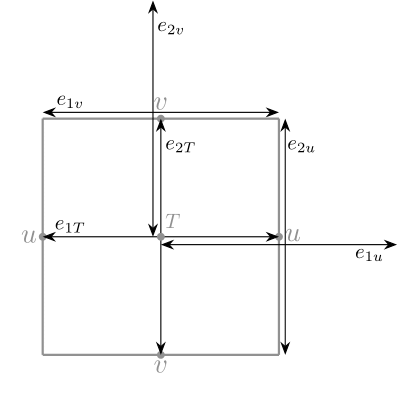}
  \end{center}
  \caption{The diffusivities $\bm{\kappa}_{11}$ and $\bm{\kappa}_{22}$ are, respectively, located at a $u$-point and $v$-point on a staggered Arakawa C-grid. 
  The normalization factors $\bm{\gamma}$ are located at $T$-points.
  Here, we scale $\bm{\kappa}_{11}$ and $\bm{\kappa}_
  {22}$ by the ratios $\mathbf{e}_{2u}/\mathbf{e}_{1u}$ and $\mathbf{e}_{1v}/\mathbf{e}_{2v}$, respectively.}
  \label{fig:alpha_def}
\end{figure}

We want to approximate a non-linear function 
\begin{align}
    f: (\bm{\kappa}_{11}, \bm{\kappa}_{22}, \mathbf{e}_{1v}, \mathbf{e}_{2v}, \mathbf{e}_{1u}, \mathbf{e}_{2u}, \mathbf{w}, \mathbf{m})\;\to \bm{\gamma} \label{eq:f}
\end{align}
where $\bm{\kappa}_{11}$ and $\bm{\kappa}_{22}$ are the diagonal elements of the horizontal diffusion tensor, with physical units of squared length; $\mathbf{e}_{1v}, \mathbf{e}_{2v}, \mathbf{e}_{1u}$ and $\mathbf{e}_{2u}$ are scale factors that define the local directional grid lengths at the $v$- and $u$-points indicated in Figure~\ref{fig:alpha_def}; $\mathbf{w}:=\mathbf{e}_{1T}\mathbf{e}_{2T}$ is the grid cell area defined at the cell center or $T$-point in Figure~\ref{fig:alpha_def}; $\mathbf{m}$ is the land-sea mask, also defined at the $T$-point; and $\bm{\gamma}$ denotes the associated normalization factor. 
Here, we are using conventional notation from the ocean model NEMO (Nucleus for European Modelling of the Ocean; \citealp{gurvan_madec_2023_8167700}).
All of these inputs are vectors of the size of the grid, some of them being located in the centers of the cells, some on the edges. 
Note that we limit ourselves to the case of diagonal diffusion tensors here, which implies that the principal directions of anisotropy are assumed to be aligned with the grid edges.
This assumption is currently used in the operational ocean DA systems at ECMWF and Met Office.
The diagonal elements of the tensor, $\bm{\kappa}_{11}$ and $\bm{\kappa}_{22}$, are hereafter called {\em diffusivity fields} or {\em diffusivities}.
The problem also depends on external parameters that we consider as constant here. 
These parameters include the total number of diffusion steps $M$ and the type of boundary condition. 
In our experiments, $M$ is set to 10 to represent an approximate Gaussian function, and a Neumann boundary condition is imposed where the normal derivative at coastlines is set to zero.

In the specific context of diffusion, we can simplify the problem by performing a local re-scaling of the diffusivity fields $\bm{\kappa}_{11}$ and $\bm{\kappa}_{22}$.
This allows us to account for the land-sea mask and for the grid structure implicitly.
More precisely, problem~\eqref{eq:f} can be transformed, without approximation, as:
\begin{align}
    \widetilde{f}: (\bm{\alpha}_1, \bm{\alpha}_2, \mathbf{w}) \; \to\; \bm{\gamma} \label{eq:ftilda}
\end{align}
where we have introduced the scaled diffusivity fields
\begin{align*}
    \bm{\alpha}_1 = \bm{\epsilon}_u\frac{\bm{\kappa}_{22}\bm{e}_{1v}}{\bm{e}_{2v}}
    \hspace{5mm} \mbox{and} \hspace{5mm}
    \bm{\alpha}_2 = \bm{\epsilon}_v\frac{\bm{\kappa}_{11}\bm{e}_{2u}}{\bm{e}_{1u}},
    \label{eq:def_alpha}
\end{align*}
$\bm{\epsilon}_u$ and $\bm{\epsilon}_v$ being defined from the land-sea mask $\mathbf{m}$ and from the boundary condition type, as explained in \citet[][their section 5.2]{mirouze_weaver_2010}. 
This re-formulation is made possible by noting that the discretized diffusion operator can be expressed in terms of $\bm{\alpha}_1$, $\bm{\alpha}_2$ and $\mathbf{w}$ only, as shown by \citet{mirouze_weaver_2010}.
In our case, we used (zero) Neumann boundary conditions, which is equivalent to setting $\bm{\epsilon}_u$ 
 (respectively, $\bm{\epsilon}_v$) equal to zero on east-west (respectively, north-south) coastlines and equal to one elsewhere. 
Other types of boundary conditions would lead to different values of the $\bm{\epsilon}_u$ and $\bm{\epsilon}_v$ fields on the coastlines, as detailed in \citet{mirouze_weaver_2010}.

We remark that there is no loss of information through this transformation, as we have
\begin{align}
    f(\bm{\kappa}_{11}, \bm{\kappa}_{22}, \mathbf{e}_{1v}, \mathbf{e}_{2v}, \mathbf{e}_{1u}, \mathbf{e}_{2u}, \mathbf{w}, \mathbf{m}) = \widetilde{f}(\bm{\alpha}_1, \bm{\alpha}_2, \mathbf{w}).
\end{align}
The problem is thus reduced from eight input fields to three. 
Importantly, both formulations of the normalization problem are equivariant to translation, a key property that we will explore further in the following subsection.

\subsection{Equivariance to translation}
\label{sec_equivariance_to_translation}

A fully convolutional neural network represents functions that are equivariant to translation \citep[][Chapter 9.4]{GoodBengCour16}.
A function $f$ is equivariant to a spatial translation $T$ if it satisfies the following identity:
\begin{align}
    f (T(\mathbf{x})) = T(f(\mathbf{x})),
    \label{eq:equivariance}
\end{align}
where $\mathbf{x}$ denotes the inputs of the function, which should match the inputs used by the corresponding network.
As the CNN is itself equivariant to translation by construction, it is intrinsically limited on how well it can fit a function that does not satisfy this property.

\begin{figure}[!h]
  \begin{center}
    \includegraphics[width=\linewidth]{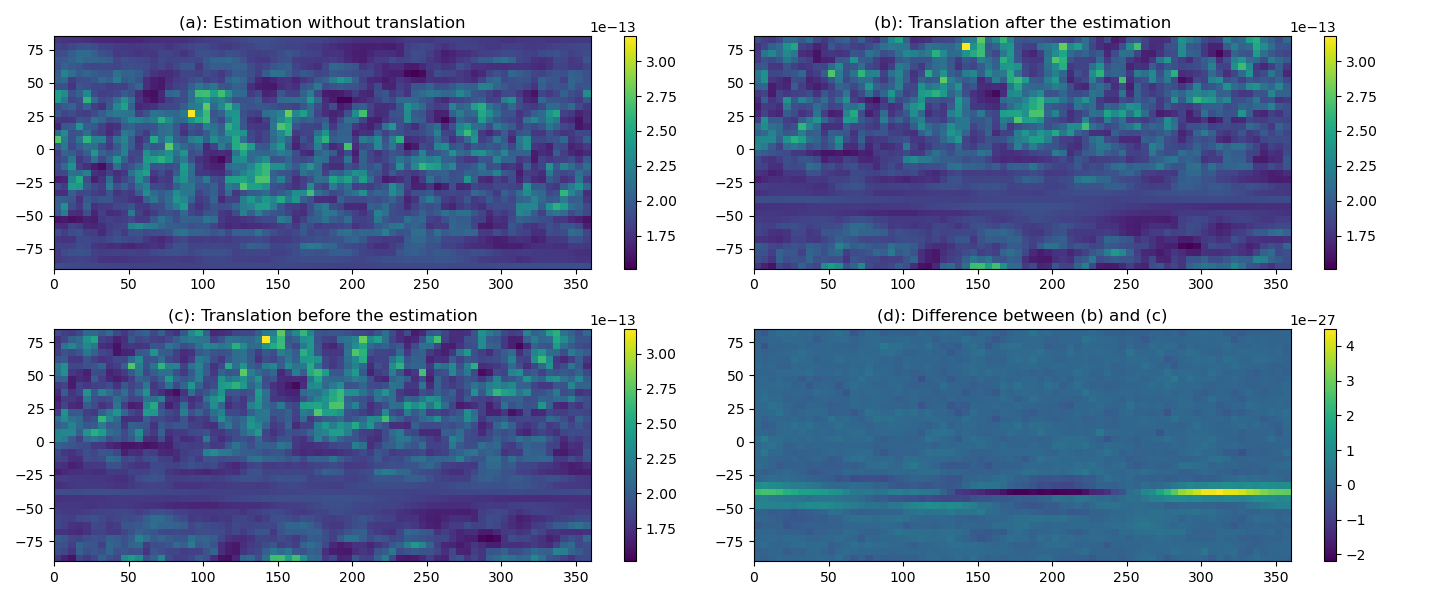 }
  \end{center}
  \caption{Numerical test illustrating the equivariance relation from Equation~\eqref{eq:equivariance}. Panel (a) shows $f(\mathbf{x})$, where $f$ is defined as in Equation~\eqref{eq:f} and $\mathbf{x}$ corresponds to random diffusivity fields $\bm{\kappa}_{11}$ and $\bm{\kappa}_{22}$, and the coefficients $\mathbf{e}_{1v}, \mathbf{e}_{2v}, \mathbf{e}_{1u}, \mathbf{e}_{2u}, \mathbf{w}$ and $\mathbf{m}$ are associated with the regular grid used with this test problem. Panel~(b) shows $T(f(\mathbf{x}))$, panel (c) shows $f(T(\mathbf{x}))$, and panel~(d) shows the relative difference between (b) and (c) (\textit{i.e.}, $| 1 -f(T(\mathbf{x}))/T(f(\mathbf{x}))|$ ). The relative difference (d) is equal to zero at machine precision.}
  \label{fig:equiv}
\end{figure}

Both formulations \eqref{eq:f} and \eqref{eq:ftilda} of the normalization problem are equivariant to translation.
The equivariance to translation of formulation \eqref{eq:ftilda} is demonstrated in Appendix~\ref{app_equiv}. For the sake of simplicity, the analysis is restricted to one-dimensional (1D) diffusion, although the same line of reasoning can be applied to 2D diffusion. 
The equivariance to translation property is illustrated numerically in Figure~\ref{fig:equiv}.
This figure has been obtained using a 2D diffusion model in Python that closely reproduces the 2D diffusion on the sphere from NEMOVAR but uses a regular latitude/longitude grid.

It is important to remark that the equivariance to translation property would not be verified if we neglected to provide the grid information to the neural network, assuming that it could learn them, as in
\begin{align}
    f_{(\mathbf{e}_{1v}, \mathbf{e}_{2v}, \mathbf{e}_{1u}, \mathbf{e}_{2u}, \mathbf{w})}: (\bm{\kappa}_{11}, \bm{\kappa}_{22}, \mathbf{m})\; & \to \bm{\gamma}\\
    \mbox{or} \; \;f_{(\mathbf{e}_{1v}, \mathbf{e}_{2v}, \mathbf{e}_{1u}, \mathbf{e}_{2u})}: (\bm{\kappa}_{11}, \bm{\kappa}_{22}, \mathbf{m}, \mathbf{w})\; & \to \bm{\gamma}.
\end{align}
In this case, a pure CNN could not fit the function with an arbitrarily small error.
In formulation~\eqref{eq:f}, the grid parameters are needed to preserve equivariance to translation but mostly represent redundant information, making formulation~\eqref{eq:ftilda} the more natural choice.

\subsection{Generation of the data set}
\label{subsec:generation}

There are two steps in the generation of the training, validation, and test data sets: 1) design realistic input diffusivity fields; and 2) compute the associated normalization fields. 
We use the so-called ORCA1 configuration of NEMO \citep{gurvan_madec_2023_8167700}, which refers to the standard tri-polar global grid with approximately one-degree horizontal resolution. All fields are defined in the surface level of ORCA1 in our 2D diffusion experiments. 

For the first step, we generate reference diffusivity fields and add random, spatially-correlated perturbations to them to produce a set of distinct input samples, with complex spatial variations similar to those that would be obtained from ensemble methods such as the one used by \cite{weaver2020evaluation}. 
The reference field is generated following the method described in \citet[][their Section 4.6.2 and their Table 2]{nemovar}.
The reference field is then perturbed by applying a multiplicative factor defined as a spatially correlated Gaussian random field centered on 1.0 and with standard deviation 0.15. The correlation model used to define the perturbations is the same correlation model associated with the reference diffusivity fields.
Some additional procedures are then applied to the perturbed diffusivity fields. First, we reduce the value of a diffusivity field near coastlines when the geometric mean of the implied directional length-scales at the diffusivity grid point exceeds the distance to the coastline from that grid point. 
Second, we increase the value of a diffusivity when the geometric mean of the implied directional length-scales is less than the geometric mean of the grid scale factors at the diffusivity point.
We generate a total of 190 samples, which differ from each other by the multiplicative noise factor only.  

For the second step, we calculate the normalization factors for each of the 190 input samples using the brute-force method described in Section~\ref{sec:introduction}.
This method is numerically affordable for a 2D field in our relatively coarse global ocean configuration. However, for a 3D field or with higher resolution, the method would be extremely expensive and not practical for generating large data sets.
This issue is discussed in the last section of this article.

\begin{figure}[!h]
  \begin{center}
    \includegraphics[width=\linewidth]{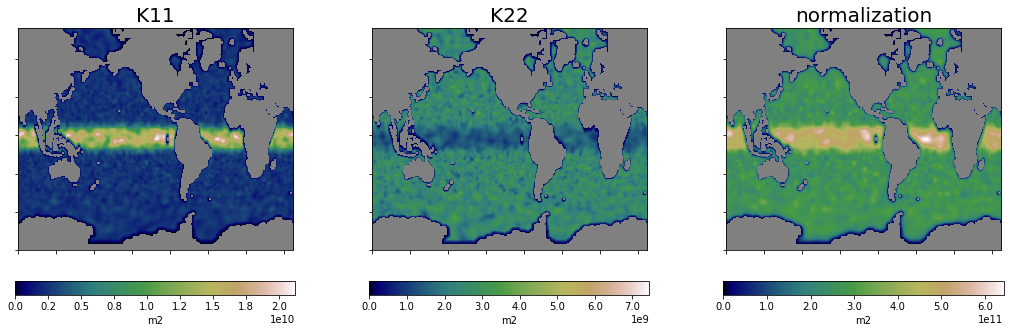}
  \end{center}
  \caption{An example of a sample of randomly-perturbed diffusivity fields $\bm{\kappa}_{11}$ and $\bm{\kappa}_{22}$ (left and middle panels) and the
 associated normalization factors (right panel). }
  \label{fig:input_output}
\end{figure}

An example of an input sample of diffusivity fields along with their corresponding true value of the normalization field is illustrated in Figure~\ref{fig:input_output}.

\subsection{Pre-processing of the data}
\label{subsec:preprocessing}

As explained in Section~\ref{sec:deep-learning}\ref{subsec:reducedim}, the diffusivity fields are scaled to reduce the number of inputs while preserving equivariance to translation.
We also observed that adding a distance-to-coast input channel reduces large errors near coastlines. The results we present here are all obtained with this additional input channel. 
The data are then separated into training, validation, and test sets containing 10, 10 and 170 samples, respectively.
We deliberately limited ourselves to 10 input/output training pairs to reflect the increased cost of data generation when considering the problem in higher dimensions or with higher resolution.
Using statistics from the training data set, each of the four input channels and the output channel are centered with respect to their global mean and normalized by their standard deviation.

Finally, the inputs are extended by periodicity in both the longitudinal and latitudinal directions by a few grid points. 
This ensures that no assumptions need to be made by the neural network near the boundaries of the ``image" while keeping the network output at the correct (original) dimensions.
This extension by periodicity accounts for the fact that the input channels are defined on different points on the staggered grid, as illustrated in Figure \ref{fig:alpha_def}.

\subsection{Model architecture and training specifications}

\begin{figure}[H]
\centering
\begin{minipage}[b]{.45\textwidth}
  \centering
  \includegraphics[width=0.7\linewidth]{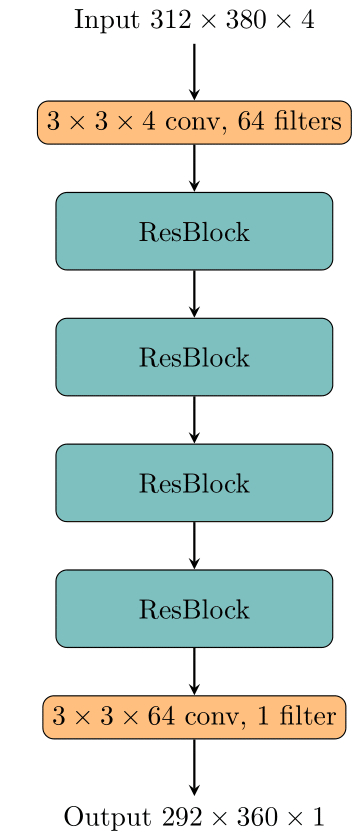}
      \captionof{figure}{Architecture containing four ResBlocks along with convolutional layers for the network input and output, marked in orange. The first layer consists of 64 filters, each convolving over the 4 input channels and making the input of a ResBlock contain 64 feature maps. Inversely, the final layer consists of only one filter to obtain the correct output dimension of $292\times 360 \times 1$.}
      \label{fig:architecture}
\end{minipage}%
\quad
\begin{minipage}[b]{.45\textwidth}
  \centering
  \includegraphics[width=0.9\linewidth]{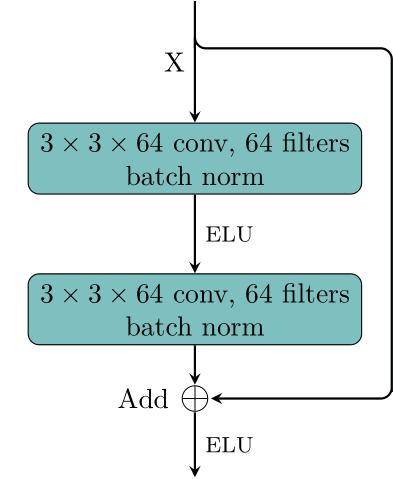}
  \captionof{figure}{Our implementation of a ResBlock consists of two 3x3x64 convolutional layers with 64 filters, each followed by batch normalization and the application of an Exponential Linear Unit (ELU) activation function. The inputs of a block are added to the output of the second convolutional layer before applying the activation function.}
  \label{fig:ResBlock}

\end{minipage}
\end{figure}

The final network, inspired by the ResNet of \cite{resnet}, consists of  Residual Blocks (ResBlocks) chained together as illustrated in Figure \ref{fig:architecture}.
A ResBlock uses skipped connections to add, or otherwise transfer, the output of one layer to a layer deeper in the network.
Our version of a ResBlock is detailed in Figure \ref{fig:ResBlock}.
The network contains 4 ResBlocks along with a layer for both the network input and output, totaling 10 convolutional layers and approximately 3 $\times 10^{5}$ trainable parameters.

We use a relative loss function where the continent grid points are masked; i.e., the model is only penalized for errors on grid points in the ocean: 
\begin{align*}
    \ell = \frac{1}{N}\sum^N_{n=0}\left(\frac{\hat{\gamma}_n^{2} - \gamma_n^{2}}{\gamma_n^{2}}\right)^2.
\end{align*}
where $\gamma_n^{2}$ is the true normalization factor, $\hat{\gamma}_n^{2}$ is the predicted normalization factor, and $N$ is the number of ocean grid points. 

The model is trained from scratch using Pytorch version 1.11.0 \citep{NEURIPS2019_9015} default weight initialization and the Adam optimizer with default learning rate of $10^{-3}$.
 
The hardware used is an Nvidia Tesla v100 16GB GPU.
The best model is selected based on performance on the validation data set after 3000~epochs, which corresponds to approximately 9~hours of training.
This high number of epochs is chosen because we use a small training data set (less than 20~MB of data), and because we observed a continued increase in accuracy both on the training and validation data sets as the number of epochs increases. 

%% file: 4-Results.tex
\section{Evaluation of the neural network}
\label{sec:results}
The neural network described in the previous section has been evaluated against the randomization method with a large sample size of $10^4$ samples. 
This sample size is a typical choice for this application, giving estimates that are considered accurate enough for operational use \citep{weaver2020evaluation}. 

\subsection{Error metrics}
We define two global error metrics to evaluate and compare different estimation methods.
Following \citet{weaver2020evaluation}, we define the relative error $\varepsilon_n$ for an ocean grid point $n$ as
\begin{align*}
    \varepsilon_n = \frac{\hat{\gamma}_n^{2} - \gamma_n^{2}}{\gamma_n^{2}},
\end{align*}
where $\gamma_n^{2}$ is the true factor computed with the brute-force method and $\hat{\gamma}_n^{2}$ is the estimate obtained using either the randomization method or the CNN. 

From this local quantity, we define two global precision measures. The first is the 
mean absolute relative error $\varepsilon$, which is defined as
\begin{align*}
    \varepsilon = \frac{1}{d}\sum^d_{l=0}\frac{1}{N}\sum^N_{n=0}|\varepsilon_n|,
\end{align*}
where $N$ is the number of ocean grid points and $d$ is the number of samples available. 
For the neural network evaluation, we present results on the test data set with {$d=170$}.
For the randomization method, we only have $d=1$. 
This is enough to have an accurate estimate, as we know the randomization errors are ``well-behaved'' in the sense that the variance estimates are unbiased and follow a $\chi^2$ distribution.

While we want the estimate to be accurate on average, we also want to control large local errors. 
We therefore introduce the mean maximum absolute relative error $\varepsilon^\infty$:
\begin{align*}
    \varepsilon^\infty = \frac{1}{d}\sum^d_{l=0}\max\limits_{0 \leq n \leq N} |\varepsilon_n|.
\end{align*}

Although the neural network is optimized to achieve a small mean square relative error through the loss function $\ell$ defined in the previous section, the final network is chosen based on the values of both $\varepsilon$ and $\varepsilon^\infty$ on the validation data.

\subsection{Results}

All neural network results shown in this section are from the test data set. 
The performance metrics of the neural network and of a randomization method with $10^4$ samples are compared in Table~\ref{tab:rel_error}. 
The neural network consistently beats the randomization method in terms of mean absolute relative error (approximately 3 times smaller). 
It has a similar performance in terms of mean maximum error. 
\begin{table}[htbp]
\centering
\begin{tabular}{||c c c ||} 
 \hline
 Method & $\varepsilon^\infty$ & $\varepsilon$ \\ [0.5ex] 
 \hline\hline
 Randomization with $10^{4}$ samples &  5.91\% & 1.15\%  \\ [1ex]
 \hline
 CNN  &  4.3\% & 0.4\%  \\ [1ex]
 \hline
\end{tabular}
\caption{Relative errors of the two estimation methods.}
\label{tab:rel_error}
\end{table}

\begin{figure}[!h]
  \begin{center}
    \includegraphics[scale=0.8]{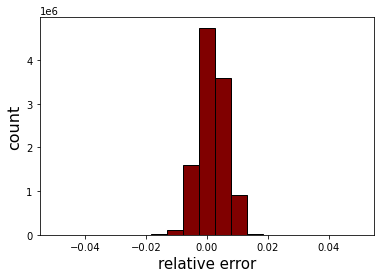}
  \end{center}
  \caption{Histogram of relative errors across all grid points and all test samples. Each bar has a width of approximately $0.53\%$.  The upper and lower bounds are set to $\pm 5\%$, leaving some extreme values hidden. These values are highlighted in Figure~\ref{fig:histo_max}. }
  \label{fig:histo_all}
\end{figure}

\begin{figure}[!h]
  \begin{center}
    \includegraphics[scale=0.8]{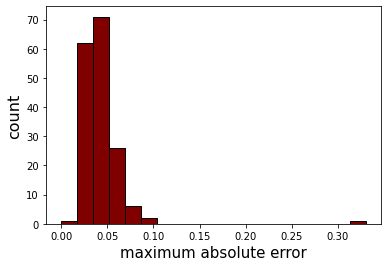}
  \end{center}
  \caption{Histogram of the maximum absolute relative error. Each bar has a width of approximately $1.73\%$. The maximum values are taken from each test sample.}
  \label{fig:histo_max}
\end{figure}

To have more information on the distribution of errors, we show the histogram of the relative errors across all grid points and all test samples in Figure~\ref{fig:histo_all}.
Similarly, Figure~\ref{fig:histo_max} shows the maximum errors of the different test samples.

Figure~\ref{fig:histo_all} shows that the neural network is globally unbiased, with errors following a Gaussian-like distribution centered on zero. 
A closer look at the distribution of the maximum absolute error in Figure~\ref{fig:histo_max} reveals an interesting outlier for one test sample, with a maximum error of $0.32$.
After investigation, this error was found to be caused by a single grid point with coastlines on several edges. In this grid cell,
the corresponding input diffusivity field $\bm{\kappa}_{11}$ has an abnormally large perturbed value, even after its original value is reduced by the distance-to-the-coast control procedure outlined in Section~\ref{sec:deep-learning}\ref{subsec:generation}.
The anomalous scaled diffusivity has a value of 10.2 after the standardization procedure explained in Section~\ref{sec:deep-learning}\ref{subsec:preprocessing}.
For comparison, the second largest scaled diffusivity in the test data set is 8.6, and the largest scaled diffusivity in the training data set is 7.5 (using the same standardization).
Importantly, this result highlights the fact that the network can make large errors for inputs that are too far from the range of values in the training data.

\begin{figure}[htbp]
  \begin{center}
    \includegraphics[scale=0.5]{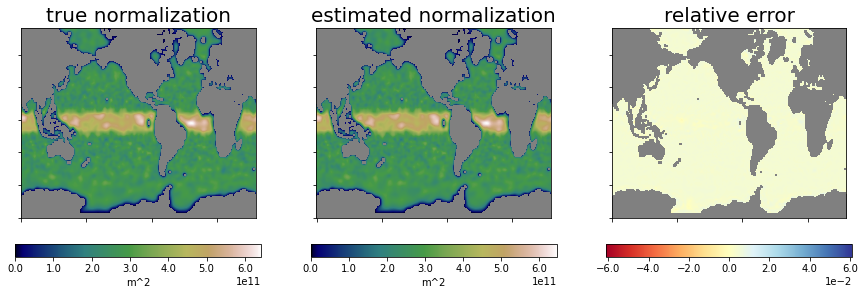}
  \end{center}
  \caption{A sample from the test data set showing the true normalization factors and the factors estimated by the network, together with their relative error.}
  \label{fig:global_error}
\end{figure}

\begin{figure}[htbp]
  \begin{center}
    \includegraphics[scale=0.5]{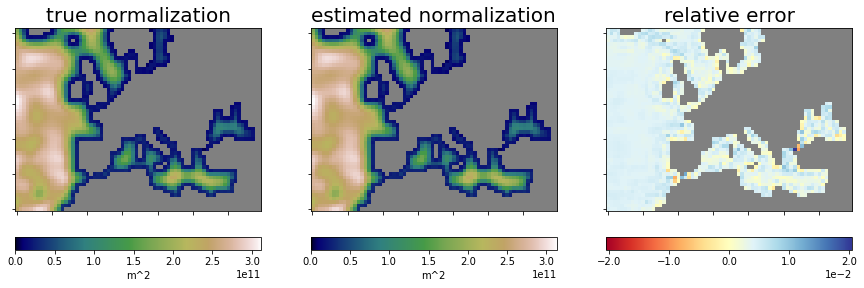}
  \end{center}
  \caption{Same as Figure~\ref{fig:global_error} but with a zoom over coastal European seas. }
  \label{fig:euro_error}
\end{figure}

An example of the neural network outputs on a random sample of the test data set is shown in Figures~\ref{fig:global_error} and \ref{fig:euro_error}.
\begin{figure}[htbp]
  \begin{center}
    \includegraphics[scale=0.45]{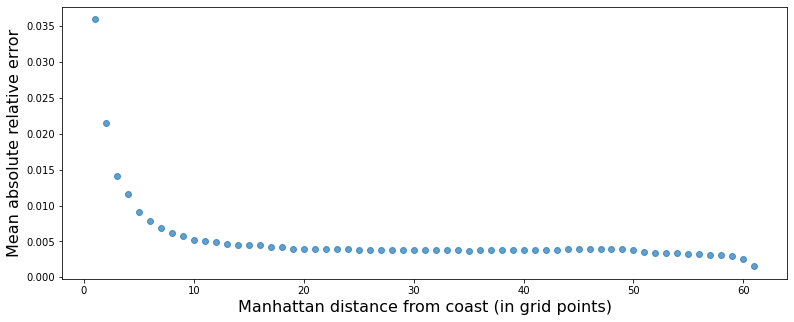}
  \end{center}
  \caption{Mean absolute relative error as a function of the Manhattan distance to the coastline. These results were obtained from a network with distance-to-coast as an additional input channel. }
  \label{fig:error_distance}
\end{figure}
There is no visible spatial structure in the errors, except when close to the coastline. 
This is evident in Figure~\ref{fig:error_distance}, which shows the error as a function of the distance to the coastline.
It clearly shows that the error is larger for points within 10 grid points of the coastlines than for points further away.
For points adjacent to a coastline, the error is up to 10 times larger.
This observation motivated the introduction of the distance-to-coast as a fourth input of the neural network, which was in fact used to produce the results in Figure~\ref{fig:error_distance}.
Although the errors are still relatively larger in the vicinity of the coastlines, this extra input channel helps maintain the amplitude of the error below that of the randomization method.

%% file: 5-Discussion.tex
\section{Summary and outlook}
\label{sec:conclusion}

In this paper, we evaluated the potential of a convolutional neural network (CNN) to estimate normalization factors for a covariance model based on a diffusion operator. 
The normalization factors are required to transform the implied diffusion-based covariance matrix into a covariance matrix with unit diagonal (i.e., a correlation matrix).
We trained a CNN on pairs of 2D diffusivity fields and the associated exact 2D normalization factors. 
The data were generated on a global grid of the NEMO ocean model \citep{gurvan_madec_2023_8167700} using a 2D diffusion operator from the NEMOVAR ocean data assimilation system \citep{nemovar}, with randomly perturbed diffusivity fields.
The trained network yields more accurate estimates than those obtained by a randomization method with 10000 samples, which is the current method of choice for this problem. 
Although it has not been quantified in this article, once trained, the CNN runs at a fraction of the computational cost of randomization.

The use of machine learning for determining normalization factors is being explored independently at the UK Met Office (James While, personal communication) and at the European Centre for Medium-Range Weather Forecasts (ECMWF; Marcin Chrust and Vincent Chabot, personal communication), using variations of the approach described in this article.
We would like to highlight three aspects of our work that we believe to be important and relevant for other machine learning approaches to this problem. \begin{enumerate}
    \item When using a CNN for this problem, the input diffusivity fields should be scaled (see Section~\ref{sec:deep-learning}\ref{sec_equivariance_to_translation}) to reduce the number of input channels, while ensuring the problem lies in the class that can be represented by the network. The appropriate scaling involves the local grid geometry and the land-sea mask used to account for the boundary conditions at coastlines.  Equivariance to translation can then be obtained by using the scaled diffusivity fields together with the local grid-cell area elements as input channels.
    \item The normalization factors are harder to determine in the vicinity of coastlines, particularly where the geometry is complex.
    In practice, adding the distance to the nearest coastline as an extra input channel to the neural network helps to reduce the estimation error. 
    The distance-to-coast input channel may become more important if the presence of a coastline outside the receptive field of the network impacts the normalization factor at a given grid point, which might occur, for instance, when moving to higher resolution grids.
    \item A surprisingly small training data set (10 pairs of 2D fields in our case) is enough to outperform the randomization method with 10000 samples.
\end{enumerate}

The network presented here is trained on a global ocean grid with coarser resolution than used operationally at ECMWF and the Met Office.
Here, the input and output fields are defined at 1~degree horizontal resolution compared to the common operational resolutions of 1/4 or 1/12~degree.
However, we expect that a CNN with similar structure would perform equally well at higher resolutions,
although the required receptive field of the neural network would possibly change due to larger diffusivities relative to the increased grid resolution.
 
In principle, the network could be extended to account for 3D diffusion or 2D$\times$1D diffusion (see Section~\ref{sec:introduction}) by using 3D convolutional kernels and by taking care to apply the same scaling procedure to the vertical diffusivities.
Within the specific 2D$\times$1D framework, which is the 3D diffusion formulation used operationally at ECMWF and the Met Office, the normalization approximation based on the separable assumption described in Section~\ref{sec:introduction} could be used with the existing 2D convolutional approach to reduce computational costs.
While the geometry of the coastlines would change in each ocean model level, the analytical problem itself would not change, and the approach we have presented should be able to account for any coastline structure. Including data from different ocean model levels in the training step would likely improve the accuracy, but would not be essential. 

Even for the 2D diffusion problem, generating a training data set of exact normalization factors would be prohibitively expensive at 1/4 and 1/12~degree resolutions. 
Indeed, the cost of a single application of the diffusion operator scales at least linearly with the number of grid points, and the brute-force approach requires as many applications as grid points. 
As a result, the brute-force approach scales at least quadratically with the number of grid points. 
Conversely, the computational cost of the randomization method only increases as the cost of a single application of the diffusion operator when increasing resolution.
The intersection point where the two methods have comparable costs is found when there are as many grid points as random samples (typically 10000); i.e., for grids even smaller than those considered in this article. 
Therefore, for high-resolution grids, the randomization approach seems to be the only viable method to generate a training data set. 
From a training point of view, it is unclear whether it is possible to beat the randomization method by training on data generated by it. 
It might be possible if the network does not try to overfit the data but manages to generalize beyond the random error in the training data. 
It might also be desirable to train on a larger random sample than is typical, to reduce sampling error as much as possible.

An alternative or complementary solution would be to train the network on a small selection of grid points only, for which the associated normalization factors 
could be estimated with the brute-force method. These points could be sampled homogeneously in the model domain, or could be sampled more heavily where the network performs more poorly; i.e., near the coastlines.
This approach could be useful at least for evaluating and testing the network with an accurate data set. 
It could also be used to provide refined training of a neural network that is pre-trained with randomization-generated data. 

Estimating normalization factors for a diffusion-based covariance operator is a computationally expensive operation, but well suited for a machine learning approach. 
Being able to estimate these factors accurately and quickly opens opportunities for using adaptive covariance parameters diffusivity fields) in operational ocean data assimilation. 
In particular, for centers that can estimate background-error correlations from an ensemble of forecasts, this would allow the use of correlations-of-the-day in parametric background-error covariance models based on the diffusion operator. 
Furthermore, this would enable the application of localization-of-the-day with ensemble-variational data assimilation (EnVar) schemes that use a diffusion operator for localizing a sample estimate of the background-error covariance matrix.

%% file: 6-Appendix.tex
\section{Equivariance to translation of one-dimensional diffusion}
\label{app_equiv}

As introduced in Section~\ref{sec:deep-learning}\ref{sec_equivariance_to_translation}, as fully convolutional networks are equivariant to translation, they are bound to be inaccurate if used to model a function that is not itself equivariant to translation. Figure~\ref{fig:equiv} illustrates experimentally that, with the right choice of inputs, the function $\widetilde{f}$ from Equation~(\ref{eq:ftilda}), is equivariant yo translation. This property can also be demonstrated analytically, which is the purpose of this appendix. 

To lighten notation, we will only consider here the case of 1D diffusion. Consequently, there is only one input feature ($\bm{\alpha}$) to consider instead of two ($\bm{\alpha}_1$ and $\bm{\alpha}_2$):
\begin{equation}
    \widetilde{f}: (\bm{\alpha}, \mathbf{w}) \; \to\; \bm{\gamma}.
\end{equation}
Our goal is to show that
\begin{equation}
\widetilde{f}(\mathbf{S}\bm{\alpha},\mathbf{S}\mathbf{w}) \, = \, \mathbf{S}\widetilde{f}(\bm{\alpha}, \mathbf{w})\, = \, \mathbf{S}\bm{\gamma},
\end{equation}
where $\mathbf{S}$ is the matrix associated with 1D translation that periodically shifts elements of a vector by one position `upwards'. 
For $p, q \in \llbracket 0,n-1\rrbracket$, the coefficient $[\mathbf{S}]_{p,q}$ on the $p$-th line and $q$-th column of $\mathbf{S}$ is defined as

\begin{equation}
[\mathbf{S}]_{p,q}=\begin{cases} 1 &\text{ if } p-q = 1;\\
                                 1 &\text{ if } p = n-1 \text{ and }  q= 0 ;\\
                                 0 &\text{ otherwise.}\\
\end{cases}
\end{equation}
The inverse translation matrix {$\mathbf{S}^{-1} = \mathbf{S}^{\rm T}$} periodically shifts elements of a vector by one position `downwards'. 

A diffusion operator $\mathbf{D}$ can be decomposed into a symmetric sequence of sub-operators ({see Equation~(4)}) from \citealp{weaver2020evaluation}):
\begin{equation}
    \mathbf{D} = \left[\mathbf{A}^{-1}\right]^{M/2} \mathbf{W}^{-1}\left[(\mathbf{A}^{-1})^{\rm T}\right]^{M/2},
    \label{eq:def_diffusion_sequence}
\end{equation}
where $\mathbf{A}$ is the self-adjoint, positive-definite matrix that must be inverted to perform one step of implicit diffusion, and $\mathbf{W}$ is a diagonal matrix whose diagonal contains the elements of the vector $\mathbf{w}$ in Equation~(\ref{eq:f}). These elements correspond to the surface of each grid cell in 2D and the length of each grid cell in 1D. The matrix $\mathbf{A}$ is self-adjoint with respect to the inner product with weighting matrix $\mathbf{W}$ (\citealp{weaver2020evaluation}, p4), which implies that the matrix $\mathbf{W}\mathbf{A}$ is symmetric; i.e., self-adjoint with respect to the canonical inner product. In terms of $\mathbf{W}\mathbf{A}$, Equation~(\ref{eq:def_diffusion_sequence}) can then be rewritten as
\begin{equation}
    \mathbf{D} = \left[\left(\mathbf{W}\mathbf{A}\right)^{-1}\mathbf{W}\right]^{M/2} \mathbf{W}^{-1}\left[\mathbf{W}\left(\mathbf{W}\mathbf{A}\right)^{-1}\right]^{M/2}.
    \label{eq:def_diffusion_sequence2}
\end{equation}

We are interested in determining the matrix expression for the `translated' diffusion operator, denoted $s(\mathbf{D})$, obtained by translating $\bm{\alpha}$ and $\mathbf{w}$. To do so, we first look for the translated counterparts of $\mathbf{W}$ and $\mathbf{WA}$, which we denote $s(\mathbf{W})$ and $s(\mathbf{WA})$, respectively. For any vector $\mathbf{v}$ of size $n$, we denote by $[\mathbf{v}]_p$ its $p$-th element with {$p \in \llbracket 0, n-1\rrbracket$}. We also assume periodic boundary conditions, which we impose by introducing cyclical indexes such that $[\mathbf{v}]_{-1} = [\mathbf{v}]_{n-1}$ and $[\mathbf{v}]_{n} = [\mathbf{v}]_{0}$.
\par
For any vector $\mathbf{x}$, we have
\begin{equation}
[\mathbf{Wx}]_p = [\mathbf{w}]_p [\mathbf{x}]_p.
\end{equation} 
Applying the translation $\mathbf{S}$ to $\mathbf{w}$ shifts periodically its elements:
\begin{equation}
[s(\mathbf{W})\, \mathbf{x}]_p = [\mathbf{w}]_{ p+1 } [\mathbf{x}]_p.
\end{equation} 
As this relation holds for all $p$, we can apply the inverse periodical shift to all indexes:
\begin{equation}
[s(\mathbf{W})\, \mathbf{x}]_{p-1} \, = \, [\mathbf{w}]_{ p } [\mathbf{x}]_{p-1}.
\end{equation} 
In matrix form, this system of equations becomes
\begin{equation}
\mathbf{S}^{-1}s(\mathbf{W})\, \mathbf{x} = \mathbf{W}\, \mathbf{S}^{-1}\, \mathbf{x}.
\end{equation} 
As this equation must hold for any vector $\mathbf{x}$, we have
\begin{equation}
s(\mathbf{W}) \, = \, \mathbf{S} \, \mathbf{W} \, \mathbf{S}^{-1}.
\label{eq:translated_W}
\end{equation} 

The same property can be derived for $s(\mathbf{WA})$. A direct relation between the elements of the matrix $\mathbf{WA}$, the scaled diffusivity vector $\bm{\alpha}$ and the vector of weighting coefficients $\mathbf{w}$ can be deduced from Equations (76) and (77) of \cite{mirouze_weaver_2010}. In a 1D domain, and with the notation used in this article, we have
\begin{align}
[(\mathbf{WA})\mathbf{x}]_p \, = \, \left( [\mathbf{w}]_p -[\bm{\alpha}]_{p-1}-[\bm{\alpha}]_{p  }\right) [\mathbf{x}]_{p}  
 -[\bm{\alpha}]_{p-1} [\mathbf{x}]_{p-1} 
 -[\bm{\alpha}]_{p  } [\mathbf{x}]_{p+1}, 
 \label{eq_system_WA}
\end{align}
for $p\in\llbracket0,n-1\rrbracket$. The scaled diffusivity $[\bm{\alpha}]_{p}$ is equal to $\kappa_p/e_p $ where $\kappa_p$ is the diffusivity on the edge between the $p$-th and {$(p+1)$}-th cells, and $e_p$ is the distance between the centres of the $p$-th and {$(p+1)$}-th cells. With $\bm{\alpha}$ and $\mathbf{w}$ translated by $\mathbf{S}$, Equation~(\ref{eq_system_WA}) becomes
\begin{align}
[s(\mathbf{WA})\mathbf{x}]_p \, = \, \left( [\mathbf{w}]_{p+1} -[\bm{\alpha}]_{p}-[\bm{\alpha}]_{p+1  }\right) [\mathbf{x}]_{p}  
 -[\bm{\alpha}]_{p} [\mathbf{x}]_{p-1} 
 -[\bm{\alpha}]_{p+1  } [\mathbf{x}]_{p+1}. 
 \label{eq_system_WA_trans}
\end{align}
As Equation~(\ref{eq_system_WA_trans}) holds for all values of $p$, we can replace $p$ by $p-1$:
\begin{align}
[s(\mathbf{WA})\mathbf{x}]_{p-1} \, = \, \left( [\mathbf{w}]_{p} -[\bm{\alpha}]_{p-1}-[\bm{\alpha}]_{p  }\right)[\mathbf{x}]_{p-1}  
 -[\bm{\alpha}]_{p-1} [\mathbf{x}]_{p-2}
 -[\bm{\alpha}]_{p  } [\mathbf{x}]_{p}, 
 \label{eq_system_WA_trans2}
\end{align}
which is equivalent to
\begin{align}
[\mathbf{S}^{-1}s(\mathbf{WA})\mathbf{x}]_{p} \, = \, \left( [\mathbf{w}]_{p} -[\bm{\alpha}]_{p-1}-[\bm{\alpha}]_{p  }\right)[\mathbf{S}^{-1}\mathbf{x}]_{p}  
 -[\bm{\alpha}]_{p-1} [\mathbf{S}^{-1}\mathbf{x}]_{p-1}
 -[\bm{\alpha}]_{p  } [\mathbf{S}^{-1}\mathbf{x}]_{p+1}.
\end{align}
In matrix form, this system can be written as
\begin{equation}
\mathbf{S}^{-1} s(\mathbf{WA})\, \mathbf{x} \, = \, \mathbf{WA} \, \mathbf{S}^{-1} \, \mathbf{x},
\end{equation}
which leads to
\begin{equation}
s(\mathbf{WA}) = \mathbf{S}\mathbf{WA}\mathbf{S}^{-1}.
\label{eq:translated_WA}
\end{equation}

By inserting Equations~(\ref{eq:translated_W}) and (\ref{eq:translated_WA}) into Equation~(\ref{eq:def_diffusion_sequence}), we obtain
\begin{equation}
    s(\mathbf{D})\, = \,\mathbf{S}\, \mathbf{D}\, \mathbf{S}^{-1}.
\end{equation}
The elements on the diagonal of $\mathbf{D}$ are the square of the inverse of the normalization factors $\gamma$:
\begin{equation}
([\bm{\gamma}]_p)^{-2} \, = \, \mathbf{e}_p^{\rm T}\, \mathbf{D}\, \mathbf{e}_p,
\end{equation}
where $\mathbf{e}_p$ is a vector whose elements are equal to zero except at position $p$ where the element is equal to 1.
We can estimate the diagonal elements of $s(\mathbf{D})$ from the inner product
\begin{align}
\mathbf{e}_p^{\rm T}\, s(\mathbf{D}) \,\mathbf{e}_p %
& \, = \, \mathbf{e}_p^{\rm T}\, \mathbf{S} \,\mathbf{D} \, \mathbf{S}^{-1} \mathbf{e}_p \nonumber\\
& \, = \, (\mathbf{S}^{-1}\mathbf{e}_p)^{\rm T}\, \mathbf{D}\, (\mathbf{S}^{-1}\mathbf{e}_p) \nonumber\\
& \, = \, (\mathbf{e}_{p+1})^{\rm T} \, \mathbf{D} \,\mathbf{e}_{p+1} \nonumber\\
& \, = \, ([\bm{\gamma}]_{p+1})^{-2} \nonumber\\
& \, = \, ([\mathbf{S}\bm{\gamma}]_{p})^{-2}, \label{eq:sD_sgamma}
\end{align}
where we have used the property that $\mathbf{S}$ is orthogonal ($\mathbf{S}^{-1} = \mathbf{S}^{\rm T}$).

Equation~\eqref{eq:sD_sgamma} shows that applying a translation $\mathbf{S}$ to the input features $\bm{\alpha}$ and $\mathbf{w}$ is equivalent to applying the same translation to $\bm{\gamma}$, and thus that $\widetilde{f}$ is equivariant to translation:
\begin{equation}
    \widetilde{f}(\mathbf{S}\bm{\alpha},\mathbf{S}\mathbf{w}) = \mathbf{S}\widetilde{f}(\bm{\alpha}, \mathbf{w}) =\mathbf{S}\bm{\gamma}.
\end{equation}
The same result can be obtained using the `downwards' translation, {$\mathbf{S}^{-1}=\mathbf{S}^{\rm T}$}, instead of the `upwards' translation, $\mathbf{S}$. By applying either of these translations multiple times, this result can be extended to any periodic translation.
Note that this property does not automatically extend to other choices of input features. For example, this result would not hold if the cell surfaces $\mathbf{w}$ were not included in the input features, or if, instead of $\bm{\alpha}$, the diffusivity field $\bm{\kappa}$ was included without the scaling factors representing the local grid geometry ($\mathbf{e}_{1v}$, $\mathbf{e}_{2v}$, $\mathbf{e}_{1u}$ and $\mathbf{e}_{2u}$ in Section~\ref{sec:deep-learning}\ref{subsec:reducedim} ).

%% file: main.bbl
\begin{thebibliography}{12}
\providecommand{\natexlab}[1]{#1}
\providecommand{\url}[1]{\texttt{#1}}
\renewcommand{\UrlFont}{\rmfamily}
\providecommand{\urlprefix}{URL }
\expandafter\ifx\csname urlstyle\endcsname\relax
  \providecommand{\doi}[1]{https://doi.org/\discretionary{}{}{}#1}\else
  \providecommand{\doi}{https://doi.org/\discretionary{}{}{}\begingroup
  \urlstyle{rm}\Url}\fi
\providecommand{\eprint}[2][]{\url{#2}}

\bibitem[{Bannister(2008)}]{bannister2008reviewb}
Bannister, R.~N., 2008: {A review of forecast error covariance statistics in
  atmospheric variational data assimilation. II: Modelling the forecast error
  covariance statistics}. \textit{Quarterly Journal of the Royal Meteorological
  Society}, \textbf{134~(637)}, 1971--1996, \doi{10.1002/qj.340}.

\bibitem[{Goodfellow et~al.(2016)Goodfellow, Bengio,, and
  Courville}]{GoodBengCour16}
Goodfellow, I.~J., Y.~Bengio, and A.~Courville, 2016: \textit{Deep Learning}.
  MIT Press, Cambridge, MA, USA, \url{http://www.deeplearningbook.org}.

\bibitem[{Guttorp and Gneiting(2006)Guttorp, and Gneiting}]{guttorp06}
Guttorp, P., and T.~Gneiting, 2006: Studies in the history of probability and
  statistics {XLIX}: On the {M}at\'ern correlation family. \textit{Biometrika},
  \textbf{93}, 989--995, \doi{10.1093/biomet/93.4.989}.

\bibitem[{He et~al.(2015)He, Zhang, Ren,, and Sun}]{resnet}
He, K., X.~Zhang, S.~Ren, and J.~Sun, 2015: Deep residual learning for image
  recognition. arXiv, \urlprefix\url{https://arxiv.org/abs/1512.03385},
  \doi{10.48550/ARXIV.1512.03385}.

\bibitem[{Lindgren et~al.(2011)Lindgren, Rue,, and
  Lindström}]{lindgren_explicit_2011}
Lindgren, F., H.~Rue, and J.~Lindström, 2011: An explicit link between
  {Gaussian} fields and {Gaussian} {Markov} random fields: the stochastic
  partial differential equation approach. \textit{Journal of the Royal
  Statistical Society: Series B (Statistical Methodology)}, \textbf{73~(4)},
  423--498, \doi{10.1111/j.1467-9868.2011.00777.x}.

\bibitem[{Madec et~al.(2023)}]{gurvan_madec_2023_8167700}
Madec, G., and Coauthors, 2023: {NEMO Ocean Engine Reference Manual}. Zenodo,
  \urlprefix\url{https://doi.org/10.5281/zenodo.8167700},
  \doi{10.5281/zenodo.8167700}.

\bibitem[{Mirouze and Weaver(2010)Mirouze, and Weaver}]{mirouze_weaver_2010}
Mirouze, I., and A.~T. Weaver, 2010: Representation of correlation functions in
  variational assimilation using an implicit diffusion operator.
  \textit{Quarterly Journal of the Royal Meteorological Society},
  \textbf{136~(651)}, 1421--1443, \doi{10.1002/qj.643}.

\bibitem[{Mogensen et~al.(2012)Mogensen, Balmaseda,, and Weaver}]{nemovar}
Mogensen, K., M.~A. Balmaseda, and A.~T. Weaver, 2012: The {NEMOVAR} ocean data
  assimilation system as implemented in the {ECMWF} ocean analysis for {System}
  4. \textbf{~(668)}, 59, \doi{10.21957/x5y9yrtm},
  \urlprefix\url{https://www.ecmwf.int/node/11174}.

\bibitem[{Paszke et~al.(2019)}]{NEURIPS2019_9015}
Paszke, A., and Coauthors, 2019: Pytorch: An imperative style, high-performance
  deep learning library. \textit{Advances in Neural Information Processing
  Systems 32}, Curran Associates, Inc., 8024--8035,
  \urlprefix\url{http://papers.neurips.cc/paper/9015-pytorch-an-imperative-style-high-performance-deep-learning-library.pdf}.

\bibitem[{Weaver and Courtier(2001)Weaver, and
  Courtier}]{weaver_correlation_2001}
Weaver, A., and P.~Courtier, 2001: Correlation modelling on the sphere using a
  generalized diffusion equation. \textit{Quarterly Journal of the Royal
  Meteorological Society}, \textbf{127~(575)}, 1815--1846,
  \doi{10.1002/qj.49712757518}.

\bibitem[{Weaver et~al.(2020)Weaver, Chrust, M{\'{e}}n{\'{e}}trier,, and
  Piacentini}]{weaver2020evaluation}
Weaver, A.~T., M.~Chrust, B.~M{\'{e}}n{\'{e}}trier, and A.~Piacentini, 2020:
  {An evaluation of methods for normalizing diffusion-based covariance
  operators in variational data assimilation}. \textit{Quarterly Journal of the
  Royal Meteorological Society}, \doi{10.1002/qj.3918}.

\bibitem[{Weaver and Mirouze(2013)Weaver, and Mirouze}]{weaver_diffusion_2013}
Weaver, A.~T., and I.~Mirouze, 2013: On the diffusion equation and its
  application to isotropic and anisotropic correlation modelling in variational
  assimilation. \textit{Quarterly Journal of the Royal Meteorological Society},
  \textbf{139~(670)}, 242--260, \doi{10.1002/qj.1955}.

\end{thebibliography}
